\catcode`\@=11\relax
\newwrite\@unused
\def\typeout#1{{\let\protect\string\immediate\write\@unused{#1}}}
\typeout{psfig: version 1.1}

%
%
\def\@nnil{\@nil}
\def\@empty{}
\def\@psdonoop#1\@@#2#3{}
\def\@psdo#1:=#2\do#3{\edef\@psdotmp{#2}\ifx\@psdotmp\@empty \else
    \expandafter\@psdoloop#2,\@nil,\@nil\@@#1{#3}\fi}
\def\@psdoloop#1,#2,#3\@@#4#5{\def#4{#1}\ifx #4\@nnil \else
       #5\def#4{#2}\ifx #4\@nnil \else#5\@ipsdoloop #3\@@#4{#5}\fi\fi}
\def\@ipsdoloop#1,#2\@@#3#4{\def#3{#1}\ifx #3\@nnil
       \let\@nextwhile=\@psdonoop \else
      #4\relax\let\@nextwhile=\@ipsdoloop\fi\@nextwhile#2\@@#3{#4}}
\def\@tpsdo#1:=#2\do#3{\xdef\@psdotmp{#2}\ifx\@psdotmp\@empty \else
    \@tpsdoloop#2\@nil\@nil\@@#1{#3}\fi}
\def\@tpsdoloop#1#2\@@#3#4{\def#3{#1}\ifx #3\@nnil
       \let\@nextwhile=\@psdonoop \else
      #4\relax\let\@nextwhile=\@tpsdoloop\fi\@nextwhile#2\@@#3{#4}}
\def\psdraft{
	\def\@psdraft{0}
}
\def\psfull{
	\def\@psdraft{100}
}
\psfull
\newif\if@prologfile
\newif\if@postlogfile
\newif\if@bbllx
\newif\if@bblly
\newif\if@bburx
\newif\if@bbury
\newif\if@height
\newif\if@width
\newif\if@rheight
\newif\if@rwidth
\newif\if@clip
\def\@p@@sclip#1{\@cliptrue}
\def\@p@@sfile#1{
		   \def\@p@sfile{#1}
}
\def\@p@@sfigure#1{\def\@p@sfile{#1}}
\def\@p@@sbbllx#1{
		\@bbllxtrue
		\dimen100=#1
		\edef\@p@sbbllx{\number\dimen100}
}
\def\@p@@sbblly#1{
		\@bbllytrue
		\dimen100=#1
		\edef\@p@sbblly{\number\dimen100}
}
\def\@p@@sbburx#1{
		\@bburxtrue
		\dimen100=#1
		\edef\@p@sbburx{\number\dimen100}
}
\def\@p@@sbbury#1{
		\@bburytrue
		\dimen100=#1
		\edef\@p@sbbury{\number\dimen100}
}
\def\@p@@sheight#1{
		\@heighttrue
		\dimen100=#1
   		\edef\@p@sheight{\number\dimen100}
}
\def\@p@@swidth#1{
		\@widthtrue
		\dimen100=#1
		\edef\@p@swidth{\number\dimen100}
}
\def\@p@@srheight#1{
		\@rheighttrue
		\dimen100=#1
		\edef\@p@srheight{\number\dimen100}
}
\def\@p@@srwidth#1{
		\@rwidthtrue
		\dimen100=#1
		\edef\@p@srwidth{\number\dimen100}
}
\def\@p@@sprolog#1{\@prologfiletrue\def\@prologfileval{#1}}
\def\@p@@spostlog#1{\@postlogfiletrue\def\@postlogfileval{#1}}
\def\@cs@name#1{\csname #1\endcsname}
\def\@setparms#1=#2,{\@cs@name{@p@@s#1}{#2}}
%
%
\def\ps@init@parms{
		\@bbllxfalse \@bbllyfalse
		\@bburxfalse \@bburyfalse
		\@heightfalse \@widthfalse
		\@rheightfalse \@rwidthfalse
		\def\@p@sbbllx{}\def\@p@sbblly{}
		\def\@p@sbburx{}\def\@p@sbbury{}
		\def\@p@sheight{}\def\@p@swidth{}
		\def\@p@srheight{}\def\@p@srwidth{}
		\def\@p@sfile{}
		\def\@p@scost{10}
		\def\@sc{}
		\@prologfilefalse
		\@postlogfilefalse
		\@clipfalse
}
%
%
\def\parse@ps@parms#1{
	 	\@psdo\@psfiga:=#1\do
		   {\expandafter\@setparms\@psfiga,}}
%
%
\newif\ifno@bb
\newif\ifnot@eof
\newread\ps@stream
\def\bb@missing{
	\typeout{psfig: searching \@p@sfile \space  for bounding box}
	\openin\ps@stream=\@p@sfile
	\no@bbtrue
	\not@eoftrue
	\catcode`\%=12
	\loop
		\read\ps@stream to \line@in
		\global\toks200=\expandafter{\line@in}
		\ifeof\ps@stream \not@eoffalse \fi
		\@bbtest{\toks200}
		\if@bbmatch\not@eoffalse\expandafter\bb@cull\the\toks200\fi
	\ifnot@eof \repeat
	\catcode`\%=14
}
\catcode`\%=12
\newif\if@bbmatch
\def\@bbtest#1{\expandafter\@a@\the#1
\long\def\@a@#1
\long\def\bb@cull#1 #2 #3 #4 #5 {
	\dimen100=#2 bp\edef\@p@sbbllx{\number\dimen100}
	\dimen100=#3 bp\edef\@p@sbblly{\number\dimen100}
	\dimen100=#4 bp\edef\@p@sbburx{\number\dimen100}
	\dimen100=#5 bp\edef\@p@sbbury{\number\dimen100}
	\no@bbfalse
}
\catcode`\%=14
\def\compute@bb{
		\no@bbfalse
		\if@bbllx \else \no@bbtrue \fi
		\if@bblly \else \no@bbtrue \fi
		\if@bburx \else \no@bbtrue \fi
		\if@bbury \else \no@bbtrue \fi
		\ifno@bb \bb@missing \fi
		\ifno@bb \typeout{FATAL ERROR: no bb supplied or found}
			\no-bb-error
		\fi
		\count203=\@p@sbburx
		\count204=\@p@sbbury
		\advance\count203 by -\@p@sbbllx
		\advance\count204 by -\@p@sbblly
		\edef\@bbw{\number\count203}
		\edef\@bbh{\number\count204}
}
%
%
\def\in@hundreds#1#2#3{\count240=#2 \count241=#3
		     \count100=\count240	
		     \divide\count100 by \count241
		     \count101=\count100
		     \multiply\count101 by \count241
		     \advance\count240 by -\count101
		     \multiply\count240 by 10
		     \count101=\count240	
		     \divide\count101 by \count241
		     \count102=\count101
		     \multiply\count102 by \count241
		     \advance\count240 by -\count102
		     \multiply\count240 by 10
		     \count102=\count240	
		     \divide\count102 by \count241
		     \count200=#1\count205=0
		     \count201=\count200
			\multiply\count201 by \count100
		 	\advance\count205 by \count201
		     \count201=\count200
			\divide\count201 by 10
			\multiply\count201 by \count101
			\advance\count205 by \count201
		     \count201=\count200
			\divide\count201 by 100
			\multiply\count201 by \count102
			\advance\count205 by \count201
		     \edef\@result{\number\count205}
}
\def\compute@wfromh{
		\in@hundreds{\@p@sheight}{\@bbw}{\@bbh}
		\edef\@p@swidth{\@result}
}
\def\compute@hfromw{
		\in@hundreds{\@p@swidth}{\@bbh}{\@bbw}
		\edef\@p@sheight{\@result}
}
\def\compute@handw{
		\if@height
			\if@width
			\else
				\compute@wfromh
			\fi
		\else
			\if@width
				\compute@hfromw
			\else
				\edef\@p@sheight{\@bbh}
				\edef\@p@swidth{\@bbw}
			\fi
		\fi
}
\def\compute@resv{
		\if@rheight \else \edef\@p@srheight{\@p@sheight} \fi
		\if@rwidth \else \edef\@p@srwidth{\@p@swidth} \fi
}
%
\def\compute@sizes{
	\compute@bb
	\compute@handw
	\compute@resv
}
%
%
\def\psfig#1{\vbox {
	%
	\ps@init@parms
	\parse@ps@parms{#1}
	\compute@sizes
	\ifnum\@p@scost<\@psdraft{
		\typeout{psfig: including \@p@sfile \space }
		\special{ps::[begin] 	\@p@swidth \space \@p@sheight \space
				\@p@sbbllx \space \@p@sbblly \space
				\@p@sbburx \space \@p@sbbury \space
				startTexFig \space }
		\if@clip{
			\typeout{(clip)}
			\special{ps:: \@p@sbbllx \space \@p@sbblly \space
				\@p@sbburx \space \@p@sbbury \space
				doclip \space }
		}\fi
		\if@prologfile
		    \special{ps: plotfile \@prologfileval \space } \fi
		\special{ps: plotfile \@p@sfile \space }
		\if@postlogfile
		    \special{ps: plotfile \@postlogfileval \space } \fi
		\special{ps::[end] endTexFig \space }
		\vbox to \@p@srheight true sp{
			\hbox to \@p@srwidth true sp{
				\hfil
			}
		\vfil
		}
	}\else{
		\vbox to \@p@srheight true sp{
		\vss
			\hbox to \@p@srwidth true sp{
				\hss
				\@p@sfile
				\hss
			}
		\vss
		}
	}\fi
}}
\catcode`\@=12\relax

\magnification = \magstep1
\hsize=7truein  \hoffset=-0.125truein
\vsize=8.5truein  \voffset=0.75truecm
\parindent=20pt
\font\ninerm=cmr9
\font\twelverm=cmr12
\font\tenrm=cmr10
\font\twelvebf=cmbx12

\def\araa{{ARA\&A}}
\def\apj{{ApJ}}
\def\apjl{{ApJ}}
\def\apjs{{ApJS}}
\def\applopt{{Appl.Optics}}
\def\apss{{Ap\&SS}}
\def\aap{{A\&A}}
\def\aapr{{A\&A~Rev.}}
\def\aaps{{A\&AS}}
\def\azh{{AZh}}
\def\baas{{BAAS}}
\def\jrasc{{JRASC}}
\def\memras{{MmRAS}}
\def\mnras{{MNRAS}}
\def\pra{{Phys.Rev.A}}
\def\prb{{Phys.Rev.B}}
\def\prc{{Phys.Rev.C}}
\def\prd{{Phys.Rev.D}}
\def\prl{{Phys.Rev.Lett}}
\def\pasp{{PASP}}
\def\pasj{{PASJ}}
\def\qjras{{QJRAS}}
\def\skytel{{S\&T}}
\def\solphys{{Solar~Phys.}}
\def\sovast{{Soviet~Ast.}}
\def\ssr{{Space~Sci.Rev.}}
\def\zap{{ZAp}}
\let\astap=\aap
\let\apjlett=\apjl
\let\apjsupp=\apjs

\def\sglbaselines{\baselineskip=22pt  \lineskip=0pt  \lineskiplimit=0pt}
\def\vs{\vskip\baselineskip}  \def\vss{\vskip 6pt}  \parskip = 0pt
\def\makeheadline{\vbox to 0pt{\vskip-30pt\line{\vbox to8.5pt{}\the
                  \headline}\vss}\nointerlineskip}
\def\toppageno{\headline={\hss\tenrm\folio\hss}}
\def\footnoterule{\kern-3pt \hrule width \hsize \kern 2.6pt \vskip 3pt}
\pretolerance=10000  \tolerance=10000  \output={\plainoutput}
\def\ts{\thinspace}
\def\cl{\centerline}
\def\ni{\noindent}  \def\nhi{\noindent \hangindent=1.0truecm}
\def\h{\hfill}
\def\0{\phantom{0}}  \def\1{\phantom{1}}  \def\d{\phantom{.}}
\let\reference=\nhi
\def\cite#1{\null}
\def\etal{et~al.}
\def\a{{\it a}}\def\b{{\it b}}\def\c{{\it c}}\def\d{{\it d}}\def\e{{\it e}}
\def\gapprox{$_>\atop{^\sim}$}  \def\lapprox{$_<\atop{^\sim}$}
\def\kms{km~s$^{-1}$}
\def\s{\ifmmode ^{\prime\prime} \else $^{\prime\prime}$ \fi}
\def\min{\ifmmode ^{\prime} \else $^{\prime}$ \fi}
\def\deg{\ifmmode ^{\circ} \else $^{\circ}$ \fi}
\def\mum{\ifmmode {\mu \rm m} \else ${\mu}$m \fi}
\def\mumd{\ifmmode {\mu \rm m.} \else ${\mu}$m. \fi}
\def\mumc{\ifmmode {\mu \rm m,} \else ${\mu}$m, \fi}
\def\Brg{\ifmmode {{\rm Br}\gamma} \else ${{\rm Br}\gamma}$ \fi}
\def\Hua{\ifmmode {{\rm Hu}\alpha} \else ${{\rm Hu}\alpha}$ \fi}
\def\Hb{\ifmmode {{\rm H}\beta} \else ${{\rm H}\beta}$ \fi}
\def\aB{\ifmmode {{\alpha}_{B}} \else ${{\alpha}_{B}}$ \fi}
\def\aHbe{\ifmmode {{\alpha}_{Hb\,{\rm eff}}}
	  \else ${{\alpha}_{\Hb\,{\rm eff}}}$ \fi}
\def\gHb{\ifmmode {{\gamma}_{Hb}}
	  \else ${{\gamma}_{\Hb}}$ \fi}
\def\gHua{\ifmmode {{\gamma}_{Hua}}
	  \else ${{\gamma}_{\Hua}}$ \fi}
\def\NeII{[Ne$\,${\ninerm II}] }
\def\HII{H$\,${\ninerm II} }
\def\ArIII{[Ar$\,${\ninerm III}] }
\def\SIV{[S$\,${\ninerm IV}] }
\newdimen\sa  \def\sd{\sa=.1em  \ifmmode $\rlap{.}$''$\kern -\sa$
                                \else \rlap{.}$''$\kern -\sa\fi}
\def\msun {M$_{\odot}$~}  \def\msund{M$_{\odot}$}  \def\mbh{$M_{\bullet}$~}
\def\Lsun{\ifmmode {{L}_{\odot}} \else ${{L}_{\odot}}$ \fi}
\def\Ll{\ifmmode {{L}_{Lyc}} \else ${{L}_{Lyc}}$ \fi}
\def\m31{M{\ts}31}  \def\mm32{M{\ts}32}  \def\mmm33{M{\ts}33}
\def\section#1#2{\vs\ni {\bf #1.}~~{\bf #2}\vs}
\def\subsection#1#2#3{\vs\cl {{\sl #1.#2.}~~{\sl #3}}\vs}
\footline={\hfil}
$\left.\right.$
\vskip1.5true in

{
\cl {\bf The 8--13 \mum Spectrum of Arp 299C}
\vs
\cl{C.~C.~Dudley and C.~G.~Wynn-Williams}
\cl{\sl Institute for Astronomy, University of Hawaii,}
\cl{\sl 2680 Woodlawn Drive, Honolulu, HI 96822}

\vskip1truein
\cl{To Appear in}

\cl{\sl The Astrophysical Journal Letters}

\vfill
\eject
\footline{\hss\tenrm\folio\hss}

\cl{\bf Abstract}
\ni{Arp 299C is a $5\times 10^{10}$ \Lsun infrared source in the merging
galaxy system Arp
299.  It is the most luminous object known that is not obviously associated
with a galaxy nucleus. Our data show that the 8--13 \mum
spectrum of Arp 299C has a strong \NeII emission line and emission
features like those of polycyclic aromatic hydrocarbon (PAH)
molecules.  These
emission features are characteristic of \HII regions associated with a burst of
star formation. We did not detect any high-excitation ionic lines
characteristic of an active galactic nucleus (AGN), nor a deep
silicate absorption feature that might
indicate a hidden compact nucleus. We deduce that Arp 299C is powered by an
intense burst of star formation.}
\vs

\ni{{\sl Subject headings:} galaxies: individual (NGC 3690/IC 694: Arp 299: Mkn
171) ---  galaxies: interaction: --- galaxies: active --- dust: unidentified
features}

\section{1}{Introduction}

The interacting galaxy system Arp 299, also known as Markarian 171 and
\hbox{NGC 3690/IC 694,}
is relatively nearby ($D=42$ Mpc, $H_0=75$ km s$^{-1}$ Mpc$^{-1}$), allowing
detailed
spatial study.  It has a large far-infrared luminosity ($ L_{\rm FIR} \approx
5\times 10^{11}$ \Lsun; Soifer \etal~1987) and a disturbed
morphology.  Of the components
identified by Gehrz, Sramek, \& Weedman (1983), component C has the
warmest near-infrared colors,
with significant nonstellar continuum emission as short as 2.3 \mum
(Ridgway \& Wynn-Williams 1993).
Components A and B
are identified with the nuclei of IC~694 and NGC 3690, respectively,
but component C lies in or is projected against the disk of NGC 3690
about 1.7 kpc from the nucleus.
Using its 25 \mum flux density as a guide, Wynn-Williams \etal~(1991)
estimate that
component C emits 10\%
($5 \times 10^{10} \Lsun$) of the total bolometric luminosity of Arp 299.
This exceeds the total far-infrared luminosity of the
Milky Way by a factor of 3.  It is also 40 times more luminous
than any of the giant \HII regions in M$\,$101.
Component C is the most luminous known region in the universe that is
not definitely
associated with a galactic nucleus.

The physical conditions in Arp 299C are also extreme.  Based on a
combination of $K$-band imaging and the CO(1--0) interferometry maps of
Sargent \& Scoville (1991), Wynn-Williams \etal~(1991) estimate that object C
might contain $7 \times 10^8$ \msun of gas within a 200 pc diameter
region.  The mean density of such a region, 6000 atoms cm$^{-3}$, would
be more than than 1000 times greater than the mean interstellar
density in the plane of the Milky Way.  Subsequently, Solomon, Downes,
\& Radford (1992) made HCN(1-0) observations of Arp 299C which confirmed the
presence of large quantities of molecular gas with densities of the
order of $10^4$ cm$^{-3}$.

The 10 \mum spectral region is very useful for determining the dominant
luminosity source of infrared galaxies.  Roche \etal~(1991) have shown
that galaxies with AGNs produce either flat
featureless spectra or deep silicate absorption, while those with
luminous \HII regions produce spectra that are dominated by emission
bands of PAH molecules (Puget \& L\'eger 1989) or other organic
compounds (Duley 1989;
Sakata \& Wada 1989).  There is a problem, though, when trying to distinguish
between silicate absorption and a PAH-dominated spectrum based on
narrowband photometry alone, since a silicate absorption feature can
be mimicked by the combination of rising thermal dust emission between
10 and 13 \mum and the long wavelength shoulder of the very strong 7.7
\mum PAH feature.  We have therefore obtained an 8--13 \mum spectrum
of Arp 299C with resolution
$\lambda/\Delta\lambda\approx 50$
order to distinguish between a silicate dominated and
a PAH dominated source.

\vskip-0.5true in

\centerline{\psfig{figure=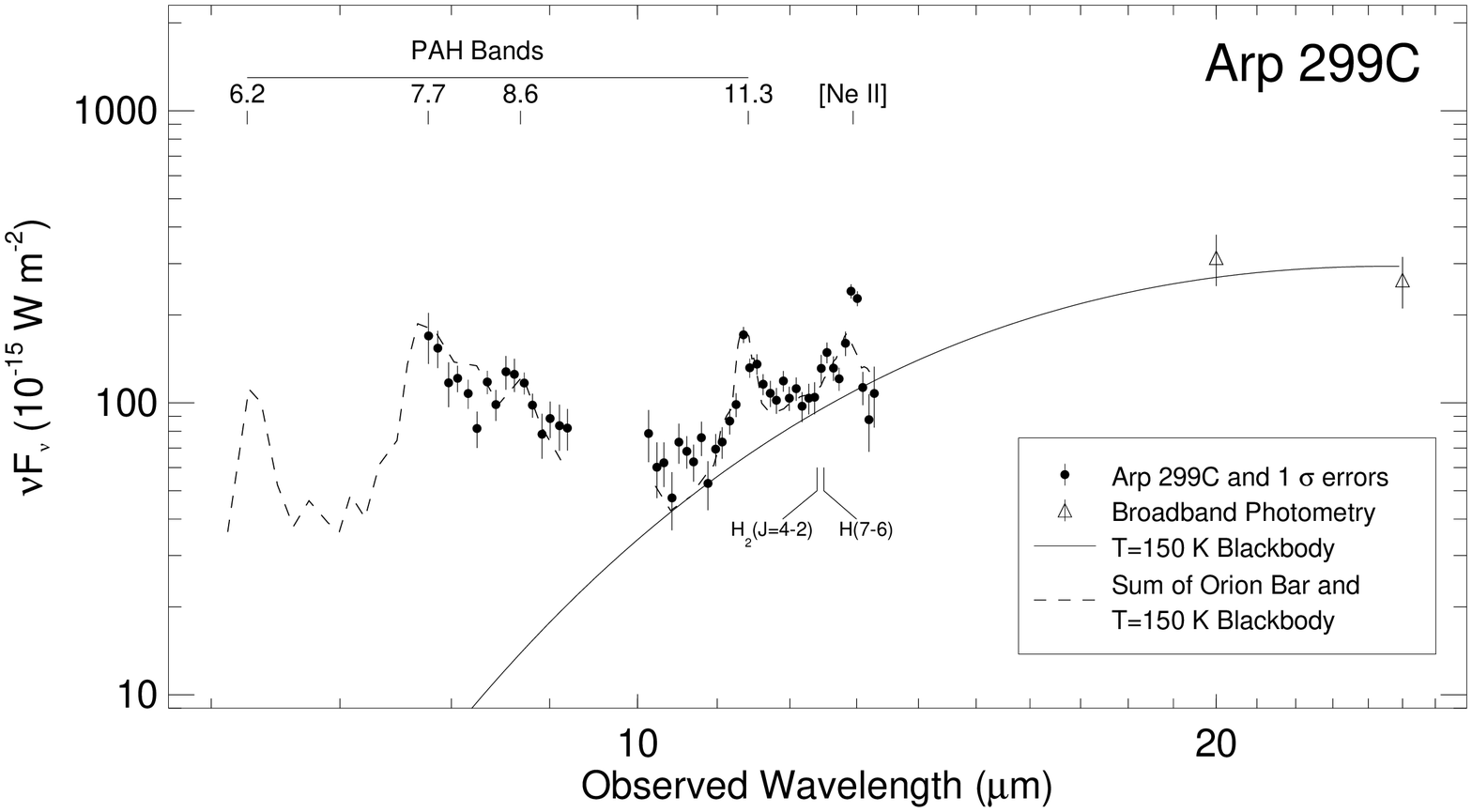,height=4.5in}}

\ni Fig. 1.  The 8--13 \mum spectrum of Arp 299C.
The triangles are from Wynn-Williams \etal~(1991).
The solid line
represents a 150 K blackbody fitted
at 13, 20, and 25 \mumd  The dashed line is a least-squares fit of the
spectrum of the Orion Bar between 10.9 and
11.7 \mumd  Note that the
\hbox{ 7.7 \mum} PAH emission dominates the emission shortward of 9
\mumc while the underlying continuum becomes increasingly
important toward longer wavelengths.  The redshifted positions of
atomic and molecular hydrogen emission lines discussed in \S 3.3 are
also indicated.

\section{2}{Observations and Data Reduction}

The data were obtained in clear weather on the night of 1992 February 16/17
at the United Kingdom Infrared
Telescope (UKIRT).\footnote{$^1$}{The
United Kingdom Infrared Telescope is operated by the Royal Observatory,
Edinburgh, for the U.~K. Science and Engineering Research Council.}
We used the 32 element 10 and 20 \mum cooled grating
spectrometer (CGS3) with a 5.5\s full width at ten percent power
circular aperture.  Our
aperture was centered on the 10.1 \mum peak position given in
Wynn-Williams \etal~(1991),
and the pointing  was maintained to within 2\s by offset guiding on a
nearby SAO star.   Two grating positions were required to sample fully
the spectrum yielding 64 data points spaced by 0.1 \mumd  The
observations consist
of 120 beam-switched pairs of 30 s each.
Since two grating
positions are required to produce a fully sampled spectrum, the
observations represent 2 hr of on-source
integration, but only \hbox{ 1 hr} per data point.  Background subtraction
was performed with a beam separation of 30\s in either the E--W or NE--SW
directions; no significant differences in the data were found in the two
chopping directions.  We have rejected data that fell outside of the
range 7.7--13.3 \mum (observed frame) and near the 10 \mum ozone
feature, because of
poor atmospheric transmission. We have also rejected two individual
subspectra that deviated by more than 5 $\sigma$ in their mean
signal from the rest of the data.
Thus the 1 $\sigma$ error bars in the spectrum
represent the error in the mean of 119 individual observations.  For
calibration purposes, we divided the spectrum by the spectrum of BS 4518
and multiplied it by a theoretical blackbody curve with $T=4400$ K
normalized at \hbox{ 10 \mum} to an assumed value of $N=0.77$ mag.  Since the
difference in
airmass between the calibration star and the source was $<$0.1,
no airmass corrections
have been applied.  We
estimate the absolute flux calibration to be better than 20\%.
When allowance is made for the very different spectral resolution
used, our measured flux densities are in satisfactory agreement with
the broadband photometry of Wynn-Williams \etal~(1991) and Gehrz
\etal~(1983).  Our wavelength calibration is based upon the spectrum
of a krypton calibration lamp.  We have not corrected the spectrum for
terrestrial or solar motion.

Our spectrum of Arp 299C (Fig.~1) reveals a  number of emission features
superimposed on a strong continuum.  The most  obvious features are the 11.3
\mum PAH band and the 12.81 \mum \NeII fine structure line. The weaker 8.6
\mum PAH band is marginally detected. We will argue in \S 3.1 that the increase
in
flux density from 9.0 to 7.7 \mum is a manifestation of the 7.7 \mum PAH
feature. In Table 1 we present the wavelengths and continuum-subtracted fluxes
of the emission features we detect and of some features we might have seen
but did not. The quoted errors are 1 $\sigma$ for detections and 3
$\sigma$ for upper limits.
Following Aitken \& Roche (1984), we have defined our continuum
relative to the PAH bands to be at 8.2 and 9.1 \mum for the 8.6 \mum
feature, and at 10.9 and 11.6 \mum for the 11.3 \mum feature in the
rest frame to allow comparison with their work.  The
reality of, and possible identification of, an unidentified 12.41 \mum feature
is discussed in \S~3.3.

\vfill\eject

\section{3}{Discussion}

\subsection{3}{1}{The PAH Bands and Possible Silicate Absorption}

The clear presence of the 12.81 \mum \NeII and  11.3 \mum PAH feature
indicates that Arp 299C contains powerful \HII regions.  Our equivalent width
for the 11.3 \mum feature is comparable to other galaxies where the feature
has been detected, although it is smaller than most (Roche \etal~1991).
The clear detection of the 11.3 \mum feature allows us to estimate how
much of the remainder of the 8--13 \mum emission from Arp 299C can be
explained by emission from other known PAH features in the bandpass.
As has been shown by Zavagno, Cox, \& Baluteau (1992) using IRAS Low
Resolution
Spectrograph data, there is an
excellent correlation
between the strength of the 11.3 \mum feature and the 7.7 + 8.6 \mum
feature  over
3 decades in feature luminosity in galactic \HII regions. Thus
the spectral contribution of the very broad 7.7 \mum PAH feature to the flux
density in the 7.7--9.0 \mum range can be predicted from the strength of the
much narrower and more easily measured 11.3 \mum feature.

We have used this relationship in Figure 1.
The Orion Bar, which lies about 2\min southeast of the Trapezium, is an
ionization front that shows a particularly rich PAH spectrum.
The most complete spectrum of the bar region is that published by
Roche (1989). It incorporates the Kuiper Airborne Observatory 6--9
\mum data of Bregman \etal~(1989) and the high spectral resolution
\hbox{ 10.7--13.2 \mum} observations of Roche, Aitken, \& Smith (1989).
This spectrum has
been fitted to the data between 10.9 and 11.7 \mum after first
subtracting an underlying
continuum that is
assumed to have the form of a 150 K blackbody.  The fit was performed by first
shifting the Orion spectrum by 3130 km s$^{-1}$ (Sargent \& Scoville 1991) and
then
degrading the spectral resolution of the Orion spectrum to that of Arp 299C by
means of a box average performed at our observed wavelengths.  The box width
was 0.179 \mumd

The excellent agreement between the observed and predicted spectrum in
the 7.7--9.0 \mum range means that it is not necessary to invoke any
silicate absorption
to explain the shape of the 8--13 \mum spectrum of Arp 299. As
is the case
for many luminous galaxies detected by IRAS, the spectrum is entirely
explicable in terms of emission by PAH grains and warm dust.
If, following Zavagno \etal~(1992), the flux of the PAH bands is given by about
15 times the flux in the
\hbox{ 11.3 \mum} feature,  then we can estimate the PAH luminosity of Arp
299C.  This
works out to about $2\times{10}^{9}\Lsun$, or about 4\% of the
estimated  bolometric luminosity of component C (Wynn-Williams \etal~1991).

\midinsert
$$\vbox{\offinterlineskip
\halign{&#  \cr
\multispan{5}\hfil Table 1: Narrow Emission Features in Arp 299C \hfil \cr
\noalign{\vskip3pt}
\noalign{\hrule}
\noalign{\vskip1pt}
\noalign{\hrule}
\noalign{\vskip3pt}

&& \hfil $\,$Rest Wavelength$\,$\hfil& \hfil Flux \hfil &
\hfil$\,$Equivalent Width\cr
\noalign{\vskip3pt}
& Feature \hfil&\hfil ($\mu$m)\hfil &\hfil $\,$($10^{-15}$ W m$^{-2}$)$\,$\hfil
&\hfil ($\mu$m)\hfil \cr
\noalign{\vskip3pt}
\noalign{\hrule}
\noalign{\vskip3pt}
&8.6 \mum \hfil&\hfil ${8.54}^{+0.09}_{-0.06}$\hfil &\hfil $1.2 \pm 0.4$\hfil
&\hfil $0.11 \pm 0.04$\hfil \cr
\noalign{\vskip11pt}
&\ArIII\hfil &\hfil 9.0\hfil &\hfil $<$ 0.6\hfil &  \cr
\noalign{\vskip11pt}
&\SIV\hfil &\hfil 10.5\hfil &\hfil $<$ 0.7\hfil &  \cr
\noalign{\vskip11pt}
&11.3 \mum \hfil&\hfil ${11.28}^{+0.03}_{-0.05}$\hfil  &\hfil $1.8 \pm 0.2$
\hfil & \hfil $0.22 \pm 0.02$ \hfil \cr
\noalign{\vskip11pt}
&?\hfil &\hfil ${12.41}^{+0.04}_{-0.07}$\hfil &\hfil $0.6 \pm 0.2$\hfil  &\hfil
$0.06 \pm 0.02$\hfil \cr
\noalign{\vskip11pt}
&\NeII \hfil&\hfil ${12.81}^{+0.01}_{-0.01}$\hfil  &\hfil $2.1 \pm 0.2$\hfil
&\hfil $0.25 \pm 0.03$\hfil \cr
\noalign{\vskip3pt}
\noalign{\hrule}
}}$$
\endinsert

\subsection{3}{2}{ Ionic lines}

In Table 1 we list three ionic forbidden lines that are diagnostic of
the level of excitation of the ionized medium in Arp 299C.  The \NeII
line is the only one detected, which argues for a low degree of
ionization.  A further check is that the ratio of \NeII to 11.3 \mum
PAH emission is about 0.9, which is quite typical of \HII region
galaxies (Roche \etal~1991).
Using the \Brg flux measured by Doyon, Puxley, \& Joseph (1992)
in a similar aperture, we find a
\NeII to \Brg ratio of
56. This ratio is within a factor of 2 of our predicted value of 87
based on  the
assumption of Case B recombination theory, an electron temperature of
10,000 K, a cosmic  Ne$\,${\ninerm II} number abundance of
$1\times10^{-4}$ relative to H$^+$ (Lacy 1982), and no collisional
de-excitation
($n \le 10^5$ cm$^{-3}$).  Given the uncertainties
involved in comparing data obtained with different instruments when there is
some extended emission in the beam, this is fair agreement.  And, given
the nondetection of the higher ionization lines of other
elements that
would betray the presence of a harder radiation field than expected in a
starburst, we can can conclude, as in the previous section, that there is
no evidence for an AGN.

In Figure 1 the rise in the
Orion Bar spectrum near the 12.81 \mum \NeII line is due to the \hbox{ 12.7
\mum} PAH
feature (Roche \etal~1989).  Because of the presence of a feature at 12.41
\mum discussed
in \S~3.3 and our
low spectral resolution, we are not able to estimate the contribution
of the \hbox{ 12.7 \mum} PAH feature to our \NeII measurement with confidence.
The fit in Figure 1 suggests that the
correct choice for the \NeII continuum may be higher than our choice, so
until this object is studied at higher spectral resolution, the
\NeII line flux in the table  could be considered an overestimate.

\vfill

\subsection{3}{3}{The Unidentified Feature at 12.41 \mum}

The apparent unresolved feature at a rest wavelength of 12.41 \mum is
a surprise.  We should caution first that it is only 3 $\sigma$ above an
uncertain continuum and thus may well be spurious.  The atmosphere has
a slight increase in opacity at this wavelength as well.  We will
consider two possible identifications, hydrogen
recombination and molecular hydrogen emission.   The feature is
coincident within the uncertainties in wavelength with the H (7--6)
hydrogen recombination line at 12.37 \mumd
The difficulty with this identification is that the measured strength
of the 12.41 \mum feature is 50 times stronger than we would predict
based on the
strength of the \Brg emission measured by Doyon \etal~(1992).  Such
an identification  would
predict 4 magnitudes of extinction at 2.16 \mum
assuming no absorption at 12 \mum and Case B recombination theory.
This in turn implies 40 magnitudes of extinction at V.  This amount
of extinction seems excessive, since the Lyman continuum luminosity
implied by this is almost 5 times larger than the bolometric
luminosity given by Wynn-Williams \etal~(1991).

Molecular hydrogen has a pure rotational \hbox{$J=4\rightarrow 2$} transition
at 12.27
\mumc and
our measured line strength would be suggestive of model 27 of Black \& Van
Dishoeck (1987) when compared with observations of Doyon (1990).
However, ihe implied recession velocity of the H$_2$ would be
6600$^{+1000}_{-1700}$ \kms, which is significantly different from the
velocity of the CO(1--0) of 3130 \kms.
This identification therefore seems unlikely, especially since the
CO(1--0) FWHM
for Arp 299C is only 50
\kms~(Sargent \& Scoville 1991).

A higher spectral resolution study  would help
to disentangle the possible 12.7 \mum PAH feature from the \NeII
feature and at the same time clarify the existence of the 12.41 feature.

\section{4}{Conclusions}

The 8--13 \mum spectrum of  Arp 299C is consistent with a luminosity
source that does not produce extremely high-energy photons.  We base
this conclusion on  the
lack of high-excitation forbidden lines and the consistency of our
\NeII and the \Brg flux of Doyon \etal~(1992).   Further, since we can
fit the
spectrum with such a simple model based on PAH molecules and thermal
dust emission without invoking silicate
absorption to account for the shape of the spectrum, a luminous AGN in
Arp 299C is not required by the data.  The similarity between the our
spectrum and
those of other starbursting galaxies implies that
the main source of luminosity is probably star formation.  Our fit to the
spectrum, while quite simplistic, explains about 90\% of the emission.
The apparent feature
at 12.41 \mum may not be real, and if it is real, probably cannot be
identified with either hydrogen recombination or molecular hydrogen emission.

\vs 

We would like to thank Drs. Tom Geballe, Martha Hanner, Dave Sanders,
Kris Sellgren,
and Alan
Tokunaga for helpful discussions, and Joel Aycock for assistance at
the telescope.
This work has been supported under the NSF Grant AST-8919563.
This work has benefited from the use of the NASA/IPAC Extragalactic
Database (NED), which is operated by the Jet Propulsion Laboratory,
Caltech, under contract with the National Aeronautics and Space
Administration.

\vfill
\eject


\ni {\bf{REFERENCES}}

\reference Aitken, D.~K., \&
Roche, P.~F.~1984, \mnras, 208, 751

\reference Black, J.~H.,
\& Van Dishoeck, E.~F.~1987, \apj, 322, 412

\reference Bregman, J.~D.,
Allamandola, L.~J., Tielens, A.~G.~G.~M., Geballe, T.~R., \& Witteborn,
F.~C.~1989, \apj, 344, 791

\reference Doyon, R., Puxley, P.~J.,
\& Joseph, R.~D.~1992, \apj, 397, 117

\reference Doyon, R.~1990, Ph.D. thesis

\reference Duley, W.~W.~1989, in IAU Symposium
135, Interstellar Dust, ed. L.~J. Allamandola \& A.~G.~G.~M. Tielens
(Dordrecht:
Kluwer), 141

\reference Gehrz,
R.~D., Sramek, R.~A., \& Weedman, D.~W.~1983, \apj, 367, 551

\reference Lacy, J.~H.~1982, in Galactic and
Extragalactic Infrared Spectroscopy, ed.~M.~F. Kessler \& \hbox{ J.~P.
Phillips} (Dordrecht: Reidel), 281

\reference Puget, L.~J., \&
L\'{e}ger, A.~1989, \araa, 27, 161

\reference Ridgway, S., \&
Wynn-Williams, C.~G.~1993, in preparation.

\reference Roche, P.~F., Aitken,
D.~K., Smith, C.~H., \& Ward, M.~J.~1991, \mnras, 248, 606

\reference Roche, P.~F.~1989, in Proc.~22
Eslab Symposium on Infrared Spectroscopy in Astronomy, ed.
B.~A.~Kaldeich, (Paris: ESA), 79

\reference Roche, P.~F., Aitken,
D.~K., \& Smith, C.~H.~1989, \mnras, 236, 485

\reference Sakata, A., \& Wada,
S.~1989, in IAU Symposium
135, Interstellar Dust, ed. L.~J. Allamandola \& A.~G.~G.~M. Tielens
(Dordrecht:
Kluwer), 191

\reference Sargent, A., \&
Scoville, N. 1991, \apj, 366, L1

\reference Soifer, B.~T., Sanders,
D.~B., Madore, B.~F., Neugebauer, G., Danielson, G.~E., Elias, J.~H.,
Lonsdale, C.~J., \& Rice, W.~L. 1987, \apj, 320, 238

\reference Solomon, P.~M., Downes,
D., \& Radford, S.~J.~E.~1992, \apjl, 387, L55

\reference Wynn-Williams, C.~G.,
Eales, S.~A., Becklin, E.~E., Hodapp, K.-W., Joseph, R.~D., McLean, I.~S.,
Simons, D.~A., \& Wright, G.~S.~1991, \apj, 377, 426

\reference Zavagno, A., Cox, P.,
\& Baluteau, J.~P.~1992, \aap, 259, 241

\vfil
\ni {\sl February 2, 1993, Honolulu \hfil}
\eject

}
\end